\documentclass[pdflatex,sn-mathphys-num]{sn-jnl}

\usepackage{graphicx}
\usepackage{multirow}
\usepackage{amsmath,amssymb,amsfonts}
\usepackage{amsthm}
\usepackage{mathrsfs}
\usepackage[title]{appendix}
\usepackage{xcolor}
\usepackage{textcomp}
\usepackage{manyfoot}
\usepackage{booktabs}
\usepackage{algorithm}
\usepackage{algorithmicx}
\usepackage{algpseudocode}
\usepackage{listings}

\theoremstyle{thmstyleone}

\theoremstyle{thmstyletwo}

\theoremstyle{thmstylethree}

\begin{document}

\title[]{Fermionic Backreaction on Quantum Spacetimes: Cosmological Implications}


\author*[1]{\fnm{Y.} \sur{Tavakoli}}\email{yaser.tavakoli@ug.edu.pl}
\author[2]{\fnm{A.} \sur{Khaleghi Ardabili}}\email{akhaleghiardabili@pennstatehealth.psu.edu}
\author[1]{\fnm{S.} \sur{Mosaddegh}}\email{mosaddegh.physics@gmail.com}

\affil*[1]{\orgdiv{Institute of Theoretical Physics and Astrophysics, Faculty of Mathematics, Physics and Informatics}, \orgname{University of Gda\'nsk}, 
\orgaddress{\street{Wita Stwosza 57}, \city{Gda\'nsk}, \postcode{80-308}, \country{Poland}}}

\affil[2]{\orgdiv{Department of Anesthesiology and Perioperative Medicine}, \orgname{Penn State Milton S. Hershey Medical Center}, \orgaddress{ \city{Hershey}, \postcode{17036, Pennsylvania},  \country{USA}}}

\abstract{
This article reviews a Hamiltonian framework for describing Dirac fermions propagating on quantum cosmological spacetimes within loop quantum cosmology. Expanding the fermionic field in spinor harmonics on a closed Friedmann--Lema\^itre--Robertson--Walker background reduces the dynamics to a collection of time-dependent Fermi oscillators, providing a Schr\"odinger-picture description of fermionic perturbations on a quantum geometry.
We discuss the emergence of dressed metrics in the test-field approximation, showing that massive fermions probe both temporal and spatial quantum-geometry corrections, whereas massless fermions, owing to conformal invariance, are affected only through a reparametrization of time. We further review the incorporation of fermionic backreaction within a Born--Oppenheimer framework, where the finite-dimensional Hilbert space of each fermionic mode gives rise to two distinct backreaction channels that naturally generate mode-dependent dressed (rainbow) metrics.
Finally, we discuss the cosmological implications of fermionic backreaction, including state-dependent modifications of the quantum bounce and the emergence of an effective cosmological constant in the semiclassical regime. These results highlight the distinctive role of fermionic matter in loop quantum cosmology and outline open directions for understanding quantum fields on quantum spacetimes.
}

\keywords{Loop quantum cosmology, Fermions, Dressed metric, Rainbow metric, Quantum bounce, Emergent cosmology.}

\maketitle

\section{Introduction}
\label{sec1}

Loop quantum gravity (LQG) provides a non-perturbative and background-independent approach to the quantization of general relativity, in which spacetime geometry acquires a fundamentally discrete structure \cite{Ashtekar:2004eh,Rovelli:2004tv,Thiemann:2007pyv}. Its symmetry-reduced realization, loop quantum cosmology (LQC), has emerged as a successful framework for investigating the quantum dynamics of the early Universe \cite{Ashtekar:2003hd,Ashtekar:2006wn,Ashtekar:2006rx,Bojowald:2001xe}. One of its most notable predictions is the resolution of the classical big-bang singularity through a quantum bounce, replacing the singular origin of the Universe with a deterministic evolution across the Planck regime \cite{Bojowald:2001xe,Ashtekar:2006uz,Ashtekar:2006wn}. Over the past two decades, LQC has developed into a mature framework with applications ranging from effective cosmological dynamics to quantum cosmological perturbation theory \cite{Ashtekar:2009mb,Dapor:2012jg, Agullo:2012fc,Agullo:2012sh, Lewandowski:2017cvz}.

A complete quantum description of the Universe, however, requires the consistent incorporation of matter fields. Considerable progress has been achieved in understanding scalar, tensor, and vector perturbations propagating on quantum cosmological backgrounds, leading to the development of the dressed-metric framework and related approaches to quantum field theory on quantum spacetimes \cite{Ashtekar:2009mb,Dapor:2012jg, Agullo:2012fc,Agullo:2012sh, Lewandowski:2017cvz}. Fermionic fields, despite being the fundamental constituents of the Standard Model and playing essential roles in particle physics and cosmology \cite{Dolgov:1989us,Kuzmin:1998kk,Parker:1971pt,Ribas:2005vr,Enqvist:2012im,Chimento:2007fx,Cortez:2016xsn}, have received comparatively less attention (see, e.g., \cite{DEath:1984gmo,DEath:1986lxx,Bojowald:2007nu,Lewandowski:2021bkt,ElizagaNavascues:2017adj,ElizagaNavascues:2018zzu,ElizagaNavascues:2019ccg,ElizagaNavascues:2019fkq,Scardua:2018omf,Tavakoli:2025pbx}). Their spinorial nature, Grassmann-valued canonical variables, and finite-dimensional Hilbert space distinguish them from bosonic fields and lead to qualitatively different features in their quantum dynamics and backreaction.

The purpose of this article is to review a Hamiltonian framework for Dirac fermions propagating on quantum cosmological spacetimes within LQC, developed in our recent work \cite{Tavakoli:2025pbx}, while placing it in a broader physical context and discussing several conceptual implications that were only briefly addressed previously. The construction is based on the canonical formulation of Dirac fields on a closed Friedmann--Lema\^itre--Robertson--Walker (FLRW) universe, expanded in spinor harmonics on the three-sphere and truncated at quadratic order \cite{DEath:1986lxx}. After quantization, each fermionic mode behaves as a time-dependent Fermi oscillator evolving with respect to the scalar-field relational clock provided by the background geometry.

Within the dressed-metric framework, fermionic perturbations propagate on an effective geometry determined by expectation values of quantum-geometric operators. Massive and massless fermions respond differently to the underlying quantum geometry: while massive modes experience corrections to both the lapse function and the scale factor, conformal invariance ensures that massless modes are affected only through a redefinition of conformal time. Going beyond the test-field approximation, fermionic backreaction can be incorporated through a Born--Oppenheimer treatment, leading naturally to mode-dependent dressed (rainbow) metrics. Owing to the finite-dimensional Hilbert space associated with each fermionic mode, only two independent backreaction channels arise, corresponding to the vacuum and pair-excited sectors, providing a distinctive feature absent in bosonic theories.

Besides reviewing the underlying Hamiltonian construction, we discuss the physical consequences of fermionic backreaction for quantum cosmology. These include state-dependent modifications of the effective quantum bounce, the emergence of a cosmological-constant-like contribution in the semiclassical regime, and the broader interpretation of these effects within the relational dynamics of LQC. We also compare the fermionic framework with previous studies of bosonic perturbations and outline several open problems that remain to be explored.

The article is organized as follows. Section \ref{sec:spinor-class} reviews the classical Hamiltonian formulation of Dirac fields on a closed FLRW spacetime and their quantization. Section \ref{sec:LQC} summarizes the loop quantization of the homogeneous background geometry. Section \ref{sec:Fermion-LQC} reviews the dressed-metric description of fermionic perturbations in the test-field approximation. Section \ref{sec:BReffects} discusses fermionic backreaction and the emergence of rainbow metrics. Section \ref{Sec:EarlyUniverse} examines the cosmological implications of these results, including modifications of the quantum bounce and the effective vacuum energy. Finally, Section \ref{sec:conclusion} summarizes the main conclusions and discusses future directions.

\section{Fermionic fields in a closed FLRW background}
\label{sec:spinor-class}

This section begins by examining the classical Hamiltonian formulation of a Dirac field propagating on a closed FLRW background. We then proceed to quantize the fermion field and analyze its spectrum within the Schr\"odinger picture.

\subsection{Hamiltonian of the two-component Dirac spinor}

We begin with the classical description of Dirac fermions on a spatially closed ($k=+1$) FLRW spacetime. Employing the two-component Weyl spinor formalism \cite{DEath:1984gmo,DEath:1986lxx,Wess:1992cp}, a Dirac spinor is represented as
\begin{equation}
\Phi = \begin{pmatrix} \phi_A \\ \bar{\chi}^{A'} \end{pmatrix},
\end{equation}
where $\phi_A$ and $\chi_A$ are left-handed Weyl spinors, and $\bar{\chi}^{A'}$ denotes the Hermitian conjugate. In terms of these spinors, the Dirac action in curved spacetime is given by
\begin{align}
I_{\rm ferm} = &-\frac{i}{2}\int d^4x\, e \left(\bar{\phi}^{A'}e^\mu_{AA'}D_\mu\phi^A + \bar{\chi}^{A'}e^\mu_{AA'}D_\mu\chi^A\right) + \text{H.c.} \nonumber \\
&-\frac{m}{\sqrt{2}}\int d^4x\, e\left(\chi_A\phi^A + \bar{\phi}^{A'}\bar{\chi}_{A'}\right),
\label{eq:actionFermion}
\end{align}
where $e^\mu_{AA'}$ represent the  
contraction of the Pauli matrices with the tetrad $e^\mu_a$,
$D_\mu$ is the spin covariant derivative compatible with the tetrad, $e$ is the tetrad determinant, and $m$ is the fermion mass.

The background geometry is $M = \mathbb{S}^3 \times \mathbb{R}$ with metric
\begin{equation}
ds^2 = -N_0^2(x^0)(dx^0)^2 + a^2(x^0)d\Omega_3^2,
\end{equation}
where $N_0$ is the lapse function, $a$ is the scale factor, and $d\Omega_3^2$ is the metric on the unit three-sphere with fiducial volume $V_0 = 16\pi^2$ \cite{Szulc:2006ep}. The physical volume is then 
\begin{equation}
  V = \ell^3 a^3 \qquad \text{with}\quad   \ell = V_0^{1/3}.
\end{equation}

Fermionic fields are expanded in spinor harmonics on $\mathbb{S}^3$, which are eigenspinors of the Dirac operator. For the left-handed spinor $\phi_A$, these satisfy
\begin{align}
-i n_{AA'} e^{BA'j} {}^{(3)}D_j \rho_B^{nq} &= \lambda_n \rho_A^{nq}, \nonumber \\
-i n_{AA'} e^{BA'j} {}^{(3)}D_j \bar{\sigma}_B^{nq} &= -\lambda_n \bar{\sigma}_A^{nq},
\end{align}
with eigenvalues $\lambda_n = n + \tfrac{3}{2}$ ($n = 0,1,2,\ldots$) and degeneracy $d_n = (n+1)(n+2)$. These harmonics form a complete orthonormal basis for spinors on $\mathbb{S}^3$.
The field expansions take the form \cite{DEath:1986lxx}
\begin{subequations}
\begin{align}
\phi_A &= \frac{a^{-3/2}}{2\pi}\sum_{npq}\alpha_n^{pq} \left(m_{np}\rho_A^{nq} + \bar{r}_{np}\bar{\sigma}_A^{nq}\right), \\
\chi_A &= \frac{a^{-3/2}}{2\pi}\sum_{npq}\beta_n^{pq} \left(s_{np}\rho_A^{nq} + \bar{t}_{np}\bar{\sigma}_A^{nq}\right),
\end{align}   
\label{eq:expansion}
\end{subequations}\\
where $n$, $p$ denote, respectively, the eigenmodes and spin projections of the fermionic field on the compact spatial manifold $\mathbb{S}^{3}$. Moreover, $m_{np}(x^0), r_{np}(x^0), s_{np}(x^0), t_{np}(x^0)$ are Grassmann-valued mode coefficients, and the factor $a^{-3/2}$ ensures canonical normalization \cite{Nelson:1978ex}.

Substituting Eqs.~\eqref{eq:expansion} into the action \eqref{eq:actionFermion} yields a decoupled sum over modes:
\begin{equation}
I_{\rm ferm} = \sum_{np} I_{np}(x, y, \bar{x}, \bar{y}),
\end{equation}
where for each $(n,p)$ pair we define Grassmann variables $(x, y)$ linearly related to the original coefficients. The mode action then becomes
\begin{align}
I_{np} = \int dt\, N_0\Big[ &\frac{i}{2N_0}(\bar{x}\dot{x} + x\dot{\bar{x}} + \bar{y}\dot{y} + y\dot{\bar{y}}) + \frac{\lambda_n}{a}(\bar{x}x + \bar{y}y) - m(yx + \bar{x}\bar{y})\Big].
\end{align}
The corresponding Hamiltonian for each mode is
\begin{equation}
H_{np} = N_0\big[\nu(\bar{x}x + \bar{y}y) + m(yx + \bar{x}\bar{y})\big],
\label{eq:hamiltonian_mode}
\end{equation}
with $\nu = \lambda_n/a$. The non-vanishing Dirac brackets are $\{x, \bar{x}\}^* = \{y, \bar{y}\}^* = -i$.

\subsection{Quantization of fermion field and spectrum}

Canonical quantization promotes the Grassmann variables to operators satisfying the anti-commutation relations $\{\hat{x}, \hat{\bar{x}}\} = \{\hat{y}, \hat{\bar{y}}\} = 1$, with all other anti-commutators vanishing. In the holomorphic representation \cite{Berezin:1966nc,Faddeev:1980be}, quantum states are represented as functions $\psi(x, y)$, evolving according to the Schr\"odinger equation
\begin{equation}
i\hbar \partial_{x^0}\psi = \sum_{np} \hat{H}_{np}\psi.
\end{equation}

In the harmonic time gauge ($N_0 = a^3$), the Hamiltonian operator for each mode takes the form
\begin{equation}
\hat{H}_{np}^{(\tau)} = \lambda_n\ell^{-2}\, V^{2/3}\big(\widehat{x\bar{x}} + \widehat{y\bar{y}}\big) + m\ell^{-3}\, V\big(\widehat{yx} + \widehat{\bar{x}\bar{y}}\big).
\end{equation}
Applying Weyl ordering \cite{Berezin:1976eg,Henneaux:1982ma}, this operator acts on state functions as
\begin{align}
\hat{H}_{np}^{(\tau)} = \lambda_n\ell^{-2}V^{2/3}\big(-1 + x\partial_x + y\partial_y\big) + m\ell^{-3}V\big(yx + \partial_x\partial_y\big).
\end{align}

For each mode, the local Hilbert space is four-dimensional, spanned by the Grassmann basis $\{1, x, y, xy\}$. Diagonalizing the Hamiltonian yields the energy spectrum
\begin{equation}
E_n = \{-w_n, \, 0, \, 0, \, +w_n\}, \qquad \text{where}\quad  w_n = \sqrt{\frac{\lambda_n^2}{\ell^4}V^{4/3} + \frac{m^2}{\ell^6}V^2}.
\label{eq:fermion_spectrum}
\end{equation}
These four energy levels correspond to the vacuum state, a single particle, a single antiparticle, and a particle--antiparticle pair, respectively. In the massless limit ($m = 0$), the Hamiltonian reduces to a diagonal form. This reflects the conformal invariance of massless Dirac fields, which prohibits gravitationally induced particle creation.

\section{Quantized  background in LQC}
\label{sec:LQC}

The fermionic degrees of freedom considered in this work propagate on a quantum FLRW spacetime described within LQC. Before introducing the fermionic sector, we briefly review the quantization of the homogeneous background following the standard LQC framework \cite{Ashtekar:2006es}. The background consists of a closed FLRW universe minimally coupled to a homogeneous massless scalar field $T$. Besides contributing to the cosmological dynamics, the scalar field evolves monotonically along classical solutions and therefore
provides a natural relational time variable with respect to which the quantum evolution of both geometry and matter can be described.

The classical dynamics are governed by the Hamiltonian constraint
\begin{equation}
H_{\rm geo}=H_T+H_{\rm grav}\approx0,
\end{equation}
where the scalar-field and gravitational contributions are given by
\begin{align}
H_T &= \frac{P_T^2}{2V},\\
H_{\rm grav} &=
-\frac{3}{8\pi G}\frac{\sqrt{p}}{\gamma^2}
\left[
\left(c-\frac{\ell_o}{a_o}\right)^2
-\frac{\gamma^2\ell_o^2}{a_o^2}
\right].
\end{align}
Here $(c,p)$ denote the Ashtekar--Barbero connection and densitized triad, satisfying the canonical Poisson bracket
$\{c,p\}=8\pi G\gamma/3$, where $\gamma$ is the
Barbero--Immirzi parameter. The physical volume is
$V=|p|^{3/2}$. The Hamiltonian constraint generates the dynamics of the homogeneous universe, while its vanishing reflects the time-reparametrization invariance of general
relativity.

To simplify the quantization, it is convenient to adopt harmonic time, corresponding to the lapse choice $N_\tau=a^3$. In this gauge the scalar-field
contribution becomes quadratic, making it particularly suitable for deparametrizing the quantum theory. The Hamiltonian constraint therefore takes
the rescaled form
\begin{equation}
H_{\rm geo}^{(\tau)} = \frac{P_T^2}{2\ell^3} + H_{\rm grav}^{(\tau)} \approx0.
\end{equation}

The quantization proceeds by treating the scalar field in the standard Schr\"odinger representation while polymer quantizing the gravitational sector. Unlike the Wheeler--DeWitt quantization, where the connection is represented by
a differential operator, LQC promotes holonomies of the connection to the fundamental quantum variables. As a consequence, the gravitational Hamiltonian becomes a finite-difference operator acting on a discrete geometry.
The quantum dynamics are encoded in the Hamiltonian constraint
\begin{equation}
\hat{H}_{\rm geo}^{(\tau)}\Psi_o(v,T)
=
\left(
\frac{\hat P_T^2}{2\ell^3}
+
\hat H_{\rm grav}^{(\tau)}
\right)
\Psi_o(v,T)
=
0,
\label{eq:constraint00}
\end{equation}
where the background wave function $\Psi_o(v,T)$ depends on the discrete volume label $v$ and the scalar-field time $T$. Physical states are those annihilated by this constraint and therefore satisfy the quantum analogue of the classical Hamiltonian constraint.

The gravitational operator $\hat H_{\rm grav}^{(\tau)}$ acts on the volume eigenstates $|v\rangle$, satisfying
\begin{equation}
\hat V|v\rangle
=
C|v|\ell_{\rm Pl}^3|v\rangle,
\end{equation}
where $C$ is a dimensionless constant.
The discrete spectrum of the volume operator reflects the underlying quantum
nature of spatial geometry in LQC. Consequently,
$\hat H_{\rm grav}^{(\tau)}$ acts as a second-order difference operator on the
volume lattice $\{|v\rangle\}$ rather than as a differential operator, providing
one of the characteristic signatures of the polymer quantization.

Since the scalar field evolves monotonically, it can be used to deparametrize the quantum theory and define relational evolution. The quantum Hamiltonian constraint can then be rewritten in the Klein--Gordon-like form
\begin{equation}
\hbar^2
\partial_T^2
\Psi_o(v,T)
=
-\Theta\Psi_o(v,T),
\end{equation}
where $\Theta$ is a positive, self-adjoint second-order difference operator acting exclusively on the geometric degrees of freedom. The positivity of $\Theta$ ensures the existence of a unique positive square root, allowing one to
restrict attention to the positive-frequency sector of the theory.
The resulting evolution equation assumes the familiar Schr\"odinger form
\begin{eqnarray}
-i\hbar
\partial_T
\Psi_o(v,T)
=
\hbar\sqrt{\Theta}\,
\Psi_o(v,T)
\equiv
\hat H_o\Psi_o(v,T),
\label{eq:geometry_evolution}
\end{eqnarray}
where the scalar field now plays the role of an internal clock and
$\hat H_o=\hbar\sqrt{\Theta}$ generates the relational evolution of the quantum
geometry. This deparametrized formulation provides the background dynamics on
which the fermionic perturbations will propagate.

The kinematical Hilbert space is
\[
\mathscr H_{\rm kin}^o
=
L^2(\bar{\mathbb R},d\mu_{\rm Bohr})
\otimes
L^2(\mathbb R,dT),
\]
where the first factor describes the polymer-quantized geometry and the second corresponds to the Schr\"odinger representation of the scalar field. The physical Hilbert space $\mathscr H_{\rm phys}^o$ consists of the
positive-frequency solutions of Eq.~\eqref{eq:geometry_evolution}, equipped with the conserved inner product
\begin{equation}
\langle\Psi_o|\Psi_o'\rangle_{\rm phys}
= \sum_{v\in\mathscr L_\varepsilon}
\overline{\Psi_o(v,T_0)}
\,\Psi_o'(v,T_0),
\end{equation}
evaluated on an arbitrary reference slice $T=T_0$. Since the inner product is independent of the choice of $T_0$, the relational evolution generated by $\hat H_o$ is unitary.

This quantization of the homogeneous geometry provides the quantum background for the dressed-metric framework developed in the following sections. Expectation values of geometric operators on the background state $\Psi_o(v,T)$ determine the effective quantum spacetime experienced by the
fermionic degrees of freedom.

\section{Fermion  on LQC background and dressed metrics}
\label{sec:Fermion-LQC}

In this section, we study the dynamics of a quantized Dirac field propagating on a loop quantum FLRW geometry. Following the dressed-metric approach originally developed for cosmological perturbations \cite{Ashtekar:2009mb}, we formulate a test-field framework in which fermionic degrees of freedom evolve on a quantum-corrected geometry while their backreaction on the background spacetime is neglected. The main objective is to identify the effective metric that captures the influence of quantum geometric fluctuations on fermionic propagation.

\subsection{Test-field approximation}
\label{subsec:test-field}

When the backreaction of the matter field on the background geometry is negligible, the total Hamiltonian constraint separates into a gravitational sector and a fermionic perturbation sector:
\begin{equation}
H_{\rm tot}=H_{\rm geo}^{(\tau)}+H_{\rm ferm}^{(\tau)}\approx0 .
\end{equation}
Here, $H_{\rm geo}^{(\tau)}$ governs the homogeneous quantum geometry, while $H_{\rm ferm}^{(\tau)}$ describes the dynamics of fermionic excitations propagating on this background. This decomposition is the starting point of the test-field approximation: the geometry determines the evolution of the fermions, but the fermions do not modify the geometry.

After quantization, the combined physical state 
$\Psi\in\mathscr{H}_{\rm phys}^{o}\otimes\mathscr{H}_{\rm ferm}$ is required to satisfy the quantum Hamiltonian constraint
\begin{equation}
\hat{H}_{\rm tot}\Psi
=
\Big(
\hat{H}_{\rm geo}^{(\tau)}\otimes\mathbb{I}_{\rm ferm}
+
\sum_{n,p}\hat{H}_{n,p}^{(\tau)}
\Big)\Psi=0.
\label{eq:full-quantum-constraint}
\end{equation}
Finding an exact solution to Eq.~\eqref{eq:full-quantum-constraint} is generally intractable due to the dynamic entanglement between quantum geometry and the fermionic degrees of freedom. To derive an effective description of the fermionic sector, we adopt a Born--Oppenheimer-type approximation: the background geometry acts as the heavy degree of freedom, while the fermionic perturbations are treated as light quantum fields evolving on it. Note, however, that we are still in a test-field limit, assuming the backreaction of the fermionic modes on the background geometry are negligible.

Accordingly, we assume that the physical state can be approximated by the factorized form
\begin{equation}
\Psi(v,T,\psi)
=
\Psi_o(v,T)
\otimes
\prod_{n,p}\psi_{n,p}(T,\mathbf{x}),
\label{eq:BO-ansatz}
\end{equation}
where $\Psi_o(v,T)$ is a normalized physical state of the homogeneous quantum geometry and $\psi_{n,p}$ represents the fermionic mode functions. Although this factorization neglects geometry-matter entanglement beyond leading order, it retains the quantum fluctuations of the background through expectation values evaluated in $\Psi_o$.
Substituting Eq.~\eqref{eq:BO-ansatz} into the full constraint and using the background evolution equation
\begin{equation}
-i\hbar\partial_T\Psi_o
=
\hat{H}_o\Psi_o
\equiv
\hbar\sqrt{\Theta}\Psi_o ,
\end{equation}
we project the resulting equation onto the background state by taking the inner product over the gravitational degrees of freedom. This procedure removes the background dynamics while leaving an effective evolution equation for the fermionic modes.

To ensure that the resulting fermionic Hamiltonian is compatible with the physical inner product of the background Hilbert space and remains self-adjoint, we introduce the symmetric ordering prescription
\begin{equation}
\hat{H}_{n,p}^{(T)}
:=
\ell^3
\hat{H}_o^{-1/2}
\hat{H}_{n,p}^{(\tau)}
\hat{H}_o^{-1/2}.
\label{eq:symmetric-ordering}
\end{equation}
The appearance of $\hat{H}_o^{-1/2}$ on both sides of the matter Hamiltonian is essential: it incorporates the quantum geometry contribution in a symmetric way and guarantees that the effective generator of evolution is defined with respect to the physical inner product of the background quantum theory.

After absorbing a state-dependent phase into the fermionic wave function, the evolution equation becomes
\begin{equation}
i\hbar\partial_T\, \psi_{n,p}
=
\big\langle
\hat{H}_{n,p}^{(T)}
\big\rangle_o\, 
\psi_{n,p},
\label{eq:fermion-effective-schrodinger}
\end{equation}
where
\[
\langle\hat{O}\rangle_o
:=
\langle\Psi_o|\hat{O}|\Psi_o\rangle_{\rm phys}
\]
denotes the expectation value in the quantum geometry state $\Psi_o$. Eq.~\eqref{eq:fermion-effective-schrodinger} is the central result of the dressed-metric construction. It shows that the fermionic modes evolve according to an effective Hamiltonian obtained by averaging the quantum geometric operators over the background state. Therefore, instead of propagating on a fixed classical FLRW spacetime, the fermions experience a dressed geometry whose properties depend on the quantum state of the universe.

\subsection{Emergent dressed metric for  fermion modes}
\label{subsec:dressed-metric-massive}

The effective Schr\"odinger equation \eqref{eq:fermion-effective-schrodinger} shows that the dynamics of fermionic modes are determined by expectation values of operators constructed from the quantum geometry. To identify the spacetime geometry encoded in these expectation values, we compare the resulting quantum Hamiltonian with the Hamiltonian of a classical Dirac field propagating on an FLRW background. This comparison allows us to extract an effective metric, usually referred to as the dressed metric, which incorporates the quantum fluctuations of the underlying loop quantum geometry.

The expectation value of the symmetrically ordered fermionic Hamiltonian takes the form
\begin{align}
\big\langle \hat{H}_{n,p}^{(T)} \big\rangle_o
= \lambda_n \langle \hat{\alpha} \rangle_o
\left( \widehat{x\bar{x}}+\widehat{y\bar{y}}
\right) + m\langle\hat{\beta}\rangle_o
\left(\widehat{yx}+\widehat{\bar{x}\bar{y}}
\right),
\label{eq:effective-fermion-H}
\end{align}
where the Grassmannian bilinears 
$\widehat{x\bar{x}},\widehat{y\bar{y}},\ldots$
are the fermionic creation and annihilation operators associated with the mode expansion on $\mathbb{S}^{3}$. The first term in Eq.~\eqref{eq:effective-fermion-H} represents the contribution from the fermionic kinetic energy. The second term originates from the fermion mass and couples the field directly to the physical volume of the universe.

The two terms involve different combinations of quantum geometric operators. These combinations are encoded in
\begin{equation}
\hat{\alpha}
:=
\ell^{-2}
\hat{H}_o^{-1/2}
\hat{V}^{2/3}
\hat{H}_o^{-1/2},
\qquad
\hat{\beta}
:=
\ell^{-3}
\hat{H}_o^{-1/2}
\hat{V}
\hat{H}_o^{-1/2}.
\label{eq:alpha-beta-operators}
\end{equation}
The operator $\hat{\alpha}$ is associated with the geometric factor controlling the propagation of the fermionic momentum modes, while $\hat{\beta}$ determines the contribution of the mass term. Thus, massive fermions probe two different aspects of quantum geometry: the spatial geometry through $\hat{\alpha}$ and the volume contribution through $\hat{\beta}$. This distinction will become important when discussing the difference between massive and massless fermionic propagation.

To determine the effective spacetime geometry, we compare Eq.~\eqref{eq:effective-fermion-H} with the standard first-order Hamiltonian of a classical Dirac field evolving on an FLRW metric,
\begin{equation}
d\bar{s}^{2}
=
-\bar{N}_{T}^{2}dT^{2}
+
\bar{a}^{2}(T)d\Omega_{3}^{2}.
\end{equation}
The coefficients multiplying the kinetic and mass terms in the classical Hamiltonian are uniquely fixed by the lapse function and the scale factor. Therefore, by matching the classical and quantum expressions, the lapse and scale factor of the effective geometry can be identified with the expectation values of the corresponding quantum geometric operators.
This matching gives the following:
\begin{align}
\bar{N}_{T}(T)
&=
\ell^{-3}
\left\langle
\hat{H}_o^{-1/2}
\hat{V}
\hat{H}_o^{-1/2}
\right\rangle_o ,
\label{eq:dressed_lapse}
\\[5pt]
\bar{a}(T)
&=
\ell^{-1}
\frac{
\left\langle
\hat{H}_o^{-1/2}
\hat{V}
\hat{H}_o^{-1/2}
\right\rangle_o
}{
\left\langle
\hat{H}_o^{-1/2}
\hat{V}^{2/3}
\hat{H}_o^{-1/2}
\right\rangle_o
}.
\label{eq:dressed_scale}
\end{align}
These relations provide the bridge between the quantum geometry and the effective classical description. The quantities $\bar{N}_T$ and $\bar{a}(T)$ are not fundamental background variables, but rather emergent quantities obtained after averaging quantum geometric operators over the chosen background state $\Psi_o$. Consequently, different quantum states of geometry can lead to different dressed metrics even when the underlying classical theory is the same.

Eqs.~\eqref{eq:dressed_lapse} and \eqref{eq:dressed_scale} define the components of the emergent dressed metric $\bar g_{ab}$. In the semiclassical regime, where the volume is large, $v\gg1$, and quantum fluctuations become negligible,
the expectation values approach their classical counterparts,
\begin{equation}
\bar N_T\rightarrow N_T,
\qquad
\bar a\rightarrow a(T).
\end{equation}
However, near the Planck regime, the expectation values contain contributions from quantum fluctuations and higher moments of the volume operator. Therefore, the dressed metric differs from the classical FLRW geometry and provides a quantum-corrected description of the spacetime experienced by the fermionic modes. In particular, massive fermions, which depend on both $\langle\hat{\alpha}\rangle_o$ and $\langle\hat{\beta}\rangle_o$, are sensitive to a richer set of quantum geometric information and experience a smoothed quantum bounce trajectory influenced by higher-order moments of the background state.

A particularly important simplification occurs for massless fermions. Setting $m=0$ in Eq.~\eqref{eq:effective-fermion-H} eliminates the contribution proportional to $\langle\hat{\beta}\rangle_o$. Consequently, the fermionic dynamics depend only on the single geometric quantity associated with $\hat{\alpha}$:
\begin{equation}
\frac{\bar{N}_{T}}{\bar{a}} = \ell^{-2}
\left\langle \hat{H}_o^{-1/2} \hat{V}^{2/3}
\hat{H}_o^{-1/2} \right\rangle_o
= \langle\hat{\alpha}\rangle_o .
\label{eq:ratio-lapse-scale}
\end{equation}
Eq.~\eqref{eq:ratio-lapse-scale} shows that massless fermions do not distinguish separately between the lapse and the scale factor. Instead, they only couple to the conformal structure of the dressed geometry through the ratio $\bar N_T/\bar a$. This is the quantum analogue of the conformal invariance of the classical massless Dirac equation.

Introducing the dressed conformal time variable $\bar{\eta}$,
\begin{equation}
d\bar{\eta}
:=
\frac{\bar N_T}{\bar a}dT
=
\langle\hat{\alpha}\rangle_o\,dT ,
\label{eq:dressed-conformal-time}
\end{equation}
the metric experienced by the massless fermionic modes becomes
\begin{equation}
d\bar{s}^{2}
=
\bar a^{2}(\bar{\eta})
\left(
-d\bar{\eta}^{2}
+d\Omega_{3}^{2}
\right).
\end{equation}
The conformal factor $\bar a(\bar\eta)$ does not affect the propagation of massless Dirac fields because it can be removed by the standard conformal rescaling of the fermionic variables. Therefore, massless fermions are insensitive to the details of the quantum-corrected scale factor and only experience a modification of the conformal time evolution. As a consequence, their vacuum state is preserved through the quantum bounce, leading to the suppression of gravitational particle production and ensuring adiabatic stability even in the high-curvature Planck regime.

\section{Backreaction effects and rainbow metrics}
\label{sec:BReffects}

The dressed-metric construction developed in the previous section describes fermionic fields as test fields propagating on a quantum-corrected geometry. This approximation is appropriate when the energy carried by the fermionic excitations is sufficiently small compared with the background energy scale. However, near the Planckian bounce, highly excited fermionic modes can carry a significant amount of energy and their contribution to the total Hamiltonian constraint can no longer be neglected.

In this section, we extend the Born--Oppenheimer framework to incorporate fermionic backreaction. The main consequence of this extension is that the quantum geometry itself becomes dependent on the fermionic excitation mode. As a result, different fermionic modes propagate on different effective geometries, leading to an emergent quantum-gravity realization of a \emph{rainbow metric} \cite{Dapor:2012jg,Parvizi:2021ekr}.

\subsection{Born-Oppenheimer treatment}

To include backreaction, we exploit the separation between the slow evolution of the homogeneous quantum geometry and the fast oscillatory dynamics of fermionic degrees of freedom. The key modification with respect to the test-field approximation is that the background quantum state is no longer independent of the fermionic field sector. Instead, the geometry is allowed to adjust to the energy carried by each fermionic mode.

For a given mode $(n,p)$, we therefore consider the extended Born--Oppenheimer decomposition
\begin{equation}
\tilde{\Psi}_{n,p}(v,T;x,y)
=
\tilde{\Psi}_{n,p}^{o}(v,T)
\otimes
\psi_{n,p}(T,x,y),
\label{eq:backreaction-ansatz}
\end{equation}
where $\tilde{\Psi}_{n,p}^{o}(v,T)$ denotes the quantum geometry state modified by the energy contribution of the fermionic mode $(n,p)$. Unlike the test-field state $\Psi_o$, this state contains information about the matter excitation and therefore depends explicitly on the fermionic sector.

The evolution of the coupled system is governed by the quantum Hamiltonian constraint
\begin{equation}
\hat{H}_{\rm tot}\tilde{\Psi}_{n,p}
=
\left[
\left(
-\hbar^2\partial_T^2+\Theta
\right)
\otimes\mathbb{I}_{\rm ferm}
+
2\ell^3\hat{H}_{n,p}^{(\tau)}
\right]
\left(
\tilde{\Psi}_{n,p}^{o}
\otimes
\psi_{n,p}
\right)
=0 .
\label{eq:full-BR-constraint}
\end{equation}
The first term in this equation describes the unperturbed quantum geometry, while the second term represents the energy carried by the fermionic mode. The factorization ansatz allows us to separate the two sectors and derive an effective equation for the geometry by projecting onto the fermionic state.

Taking the inner product with $\langle\psi_{n,p}|$ over the fermionic Hilbert space $\mathscr{H}_{\rm ferm}$ gives the effective equation for the background state. In this step we neglect non-adiabatic transitions between different fermionic states and assume that the time derivatives of the fermionic wave function are small compared with its rapidly varying dynamical phase. This is the standard adiabatic approximation underlying the Born--Oppenheimer separation.
Under this assumption, Eq.~\eqref{eq:full-BR-constraint} reduces to
\begin{equation}
\left(
-\hbar^2\partial_T^2+\Theta
\right)
\tilde{\Psi}_{n,p}^{o}(v,T)
+
2\ell^3
\langle\psi_{n,p}|
\hat{H}_{n,p}^{(\tau)}
|\psi_{n,p}\rangle_{\rm ferm}
\tilde{\Psi}_{n,p}^{o}(v,T)
=0 .
\label{eq:projected-geo-constraint}
\end{equation}
The additional term in Eq.~\eqref{eq:projected-geo-constraint} acts as an effective source for the gravitational sector. It represents the energy stored in the fermionic mode and modifies the quantum evolution of the geometry. Defining
\begin{equation}
\langle\hat{H}_{n,p}^{(\tau)}\rangle_{\psi}
:=
\langle\psi_{n,p}|
\hat{H}_{n,p}^{(\tau)}
|\psi_{n,p}\rangle_{\rm ferm},
\end{equation}
this contribution can be evaluated using the spectral decomposition of the Dirac operator on $\mathbb{S}^{3}$.

The localized mode Hamiltonian has the eigenvalues
\begin{equation}
\langle\hat{H}_{n,p}^{(\tau)}\rangle_{\psi}
=
\pm
\sqrt{
\frac{\lambda_n^2}{\ell^4}\hat{V}^{4/3}
+
\frac{m^2}{\ell^6}\hat{V}^{2}
}\, =\, 
\pm\hat{w}_n.
\label{eq:fermion_spectrum}
\end{equation}
The two signs correspond to the two possible fermionic sectors: the positive branch describes particle-pair excitations, while the negative branch corresponds to the filled Dirac sea contribution. Therefore, unlike bosonic excitations, fermionic backreaction is naturally restricted by the finite occupation structure of the theory.

Applying the symmetric ordering prescription introduced in Eq.~\eqref{eq:symmetric-ordering}, the effective energy contribution acting on the geometric sector is encoded in the operator
\begin{equation}
\hat{E}_n
:=
\left\langle
\hat{H}_o^{-1/2}
\hat{w}_n
\hat{H}_o^{-1/2}
\right\rangle_{\rm ferm}.
\end{equation}
This quantity contains both the fermionic excitation energy and its coupling to the quantum geometry. Substituting it into Eq.~\eqref{eq:projected-geo-constraint}, the backreacted geometric state satisfies
\begin{equation}
-\partial_T^2\tilde{\Psi}_{n}^{o}
=
(\Theta+\Theta_n)\tilde{\Psi}_{n}^{o},
\qquad \text{where} \quad 
\Theta_n
\equiv
-\frac{2\ell^3}{\hbar^2}\hat{E}_n .
\end{equation}
Thus, fermionic backreaction appears as a mode-dependent perturbation of the gravitational evolution operator. Each fermionic excitation modifies the quantum geometry through a different correction $\Theta_n$, which is the origin of the subsequent rainbow structure.
\medskip 

\noindent {\bf Remark.} The Born--Oppenheimer approximation is expected to remain valid as long as the quantum geometry evolves on a time scale much longer than that of the fermionic modes, so that non-adiabatic transitions between different background states remain suppressed. This assumption is analogous to the dressed-metric treatment of scalar perturbations in LQC and is expected to hold for semiclassical background states sufficiently away from regimes where strong mode mixing becomes important.

\subsection{Spectral perturbation and rainbow metric emergence}
\label{subsec:rainbow-metric}

The operator $\Theta_n$ modifies both the eigenvalues and eigenstates of the background quantum geometry. To determine this modification, let $|e_k\rangle$ be the eigenbasis of the unperturbed operator,
\begin{align}
\Theta|e_k\rangle=\omega_k^2|e_k\rangle .
\end{align}
Treating $\Theta_n$ as a perturbation, standard non-degenerate quantum perturbation theory gives
\begin{align}
\big(\omega_k^{(n)}\big)^2
&=\, 
\omega_k^2
+
\langle e_k|\Theta_n|e_k\rangle
+
\mathcal{O}(\Theta_n^2),
\label{eq:eigenvalues-perturbed}
\\
|e_k^{(n)}\rangle
&=\,
\underline{N}_k
\left(
|e_k\rangle
+
|\delta e_k^{(n)}\rangle
\right),
\label{eq:eigenstates-perturbed}
\end{align}
where $\underline{N}_k$ is a normalization factor.
These expressions show that backreaction affects the geometry in two ways: it shifts the spectrum of the gravitational evolution operator and modifies the corresponding quantum states. Consequently, the expectation values of geometric observables become dependent on the fermionic mode that generated the perturbation.

Evaluating the dressed fermionic Hamiltonian in the backreacted state $\tilde{\Psi}_{n,p}^{o}$ gives modified geometric expectation values:
\begin{equation}
i\hbar\partial_T\psi_{n,p}
=
\left[ \lambda_n
\big(\langle\hat{\alpha}\rangle_o
+
\langle\hat{\alpha}\rangle_n
\big)
(\widehat{x\bar{x}}+\widehat{y\bar{y}})
+
m
\left(
\langle\hat{\beta}\rangle_o
+
\langle\hat{\beta}\rangle_n
\right)
(\widehat{yx}+\widehat{\bar{x}\bar{y}})
\right]
\psi_{n,p},
\end{equation}
where
\begin{equation}
\langle\hat{\alpha}\rangle_n
:=
\langle\tilde{\Psi}_{n,p}^{o}|
\hat{\alpha}
|\tilde{\Psi}_{n,p}^{o}\rangle
-
\langle\Psi_o|
\hat{\alpha}
|\Psi_o\rangle ,
\end{equation}
and similarly for $\langle\hat{\beta}\rangle_n$.

The corrections $\langle\hat{\alpha}\rangle_n$ and $\langle\hat{\beta}\rangle_n$ encode the modification of the effective geometry produced by the fermionic energy density. Since they depend on the mode label $n$, the effective spacetime is no longer universal: different fermionic modes probe different dressed geometries.
The corresponding mode-dependent line element can therefore be written as
\begin{equation}
d\tilde{s}_{(n)}^2
=
-\big(\tilde{N}_T^{(n)}\big)^2dT^2
+
\big(\tilde{a}^{(n)}\big)^2d\Omega_3^2 .
\end{equation}
This geometry has the form of a rainbow metric, where the spacetime structure depends on the energy scale or mode number of the probing field. The mode-dependent lapse and scale factor can be expressed as deformations of the test-field dressed quantities:
\begin{subequations}
 \begin{equation}
\tilde{N}_T^{(n)}(T)
=
\bar{N}_T(T) F_n(V),
\qquad
\tilde{a}^{(n)}(T)
=
\bar{a}(T) G_n(V),
\end{equation}
with
\begin{equation}
F_n(V)
:=
1+
\frac{\langle\hat{\beta}\rangle_n}
{\langle\hat{\beta}\rangle_o},
\qquad
G_n(V)
:=
\left(
1+
\frac{\langle\hat{\beta}\rangle_n}
{\langle\hat{\beta}\rangle_o}
\right)
\left(
1+
\frac{\langle\hat{\alpha}\rangle_n}
{\langle\hat{\alpha}\rangle_o}
\right)^{-1}.
\end{equation}   
\end{subequations}
In the limit where the fermionic energy becomes negligible,
\begin{equation}
\hat{E}_n\rightarrow0,
\end{equation}
the corrections vanish,
\begin{equation}
F_n,\, G_n\rightarrow1,
\end{equation}
and the standard dressed metric of the test-field approximation is recovered.

For massless fermions, $m=0$, the only relevant geometric correction is the one associated with $\hat{\alpha}$. The conformal factor therefore simplifies:
\begin{equation}
\frac{\tilde{N}_T^{(n)}}{\tilde{a}^{(n)}}
=
\frac{\bar{N}_T}{\bar{a}}
\left(
1+
\frac{\langle\hat{\alpha}\rangle_n}
{\langle\hat{\alpha}\rangle_o}
\right)
\equiv
\frac{\bar{N}_T}{\bar{a}}
(1+\delta_n^{(1)}).
\end{equation}
Hence, for massless modes, fermionic backreaction does not introduce a deformation of the spatial conformal geometry. Instead, it modifies the relation between the internal time $T$ and the conformal time experienced by each mode. The main physical effect of backreaction is therefore a mode-dependent renormalization of the conformal clock rate.

\subsection{Contrast with bosonic backreaction}
\label{subsec:bosonic-contrast}

It is useful to compare the structure of fermionic rainbow metrics with the corresponding situation for bosonic fields. The difference originates from the distinct structure of their Hilbert spaces and occupation numbers.
\begin{enumerate}
\item \textit{Finite versus infinite dimensionality of the state space:} 
Because of Pauli exclusion, each fermionic Fourier mode $(n,p)$ has a finite-dimensional Hilbert space with energy eigenvalues
\[\{-w_n,0,0,+w_n\}.\]
Therefore, fermionic backreaction is characterized by a finite number of physical excitation channels, corresponding essentially to the vacuum and pair-excited states. Bosonic modes, in contrast, possess an infinite-dimensional Fock space with arbitrary occupation numbers $N_k$. Their energy spectrum,
\[
(N_k+1/2)\hbar\, \omega_k(V),
\]
allows infinitely many possible backreaction strengths and therefore potentially an infinite family of rainbow geometries.

\item \textit{Bounded versus unbounded backreaction strength:}
The fermionic occupation numbers satisfy
\[
N_{n,p}\in\{0,1\},
\]
which limits the maximum contribution of each mode to the background energy. In contrast, bosonic modes admit arbitrarily large occupation numbers ($N_k\propto n$ with $n=0, 1, 2, \dots$ for each mode) and therefore do not have an analogous upper bound \cite{Parvizi:2021ekr}. Nevertheless, the overall gravitational backreaction is not determined solely by the occupation numbers; it also depends on the volume scaling of the associated energy density during cosmological evolution. As shown in the next section, fermionic and scalar backreaction exhibit qualitatively different volume dependence, leading to distinct phenomenological consequences.

\item \textit{Analytic tractability:} The discrete and bounded structure of fermionic backreaction makes the perturbative analysis based on Eqs.~\eqref{eq:eigenvalues-perturbed}--\eqref{eq:eigenstates-perturbed} considerably more controlled. In contrast, bosonic backreaction generally requires additional approximations or truncations due to the unbounded occupation number spectrum.
\end{enumerate}

Table~\ref{table:SFComp} summarizes the main differences and similarities between scalar fields and Dirac fermions propagating on quantum spacetime, highlighting their distinct kinematical structures, backreaction properties, and phenomenological consequences.

\begin{table}[t]
\centering
\sf \small\begin{tabular}{@{}lll@{}}
\toprule
\textbf{\textsf{Feature}} & \quad  \textbf{\textsf{Scalar fields}}  & \quad  \textbf{\textsf{Dirac fermions}} \\
\midrule

Hilbert space per mode & \quad  Infinite & \quad  Finite \\[3pt]

Occupation number & \quad  $0, 1, 2, \dots$ & \quad  $1,2$  \\[3pt]

Backreaction channels & \quad  Infinite & \quad  Two \\[3pt]

Rainbow metric & \quad  Yes & \quad Yes \\[3pt]

Late-time vacuum energy & \quad  Model dependent & \quad  Constant for $m\neq 0$  \\[3pt]

Conformal invariance & \quad  Only conformally coupled scalar & \quad  Massless Dirac fermions  \\

\bottomrule
\end{tabular}
\caption{Comparison between scalar and fermionic fields in LQC.}
\label{table:SFComp}
\end{table}

\section{Cosmological phenomenological implications}
\label{Sec:EarlyUniverse}

The backreaction of quantized Dirac fields on a LQC background has consequences that extend beyond the microscopic description developed in the previous sections. Once the fermionic contribution is incorporated into the effective gravitational dynamics, it can modify both the behavior of the quantum bounce and the subsequent cosmological evolution. In this section, we discuss two representative phenomenological consequences. We first show how fermionic backreaction alters the effective bounce conditions through state-dependent corrections to the Hamiltonian constraint. We then discuss the emergence of an effective cosmological constant at late times arising from the backreaction of the fermionic sector on the quantum geometry.

\subsection{Bounce asymmetry and modified critical density}
\label{subsec:bounce-asymmetry}

As shown in Section~\ref{sec:BReffects}, fermionic backreaction modifies the effective Hamiltonian governing the quantum geometry through the expectation value of the physical fermionic Hamiltonian. Since vacuum and excited fermionic states contribute with opposite signs, the effective gravitational dynamics become sensitive to the occupation of the fermionic sector. Consequently, the parameters characterizing the quantum bounce are no longer universal but depend on the quantum state of the fermions.

It is important to emphasize that this dependence should not be interpreted as a breakdown of the time-reversal symmetry of the underlying quantum theory. The fundamental quantum dynamics remain invariant under time reversal. The apparent asymmetry discussed below instead reflects the fact that different fermionic quantum states generate different effective Hamiltonians, leading to distinct relational cosmological histories.

The effective energy density associated with the $n$-th fermionic mode is
\begin{equation}
\rho_n(T)
\, =\, 
\big\langle
:\widehat{V^{-1}(T)\, E_n^{(T)}}:
\big\rangle ,
\end{equation}
where $E_n^{(T)}$ denotes the physical Hamiltonian of the $n$-th fermionic mode with respect to the relational clock $T$, and the colons denote normal ordering. This quantity represents the contribution of a single fermionic mode to the effective matter source entering the gravitational dynamics.

Near the quantum bounce, where the physical volume approaches its minimum value $V_b$, the dominant contribution to the fermionic energy density becomes
\begin{equation}
\rho_n \simeq 2\lambda_n\ell
\left\langle
\hat{H}_o^{-1/2}
\hat{V}^{-1/3}
\hat{H}_o^{-1/2}
\right\rangle.
\end{equation}
The inverse one-third power of the volume operator is a characteristic feature of the fermionic sector and reflects the way Dirac fields couple to the quantum geometry. By comparison, the backreaction energy density of a scalar field was found in Ref.~\cite{Parvizi:2021ekr} to scale as
\begin{eqnarray}
\rho^{\rm (scalar)}_k
\propto
\big\langle
\hat V^{-4/3}
\big\rangle,
\end{eqnarray}
which exhibits a considerably stronger dependence on the physical volume. Consequently, the scalar-field backreaction grows more rapidly as the universe approaches the Planck regime, whereas the fermionic contribution varies more mildly with the volume. Conversely, during the subsequent expansion the scalar backreaction is diluted much more efficiently, while the fermionic contribution decreases more slowly. This different scaling behavior implies that fermionic backreaction can remain relevant over a broader range of the cosmological evolution and can generate observable corrections to the effective Hamiltonian constraint (see, e.g., the next subsection).

The origin of these corrections can be traced to the fermionic energy spectrum. Since the energy of each mode enters with opposite signs for the vacuum and pair-excited sectors,
\begin{align}
E_n=\mp w_n,
\end{align}
their gravitational backreaction also has opposite effects. The vacuum contribution effectively increases the gravitational potential, whereas excited fermionic states reduce it. Consequently, the density at which the bounce occurs acquires a dependence on the fermionic occupation state.
Schematically, this modification can be expressed as
\begin{align}
\rho_{\rm crit}^{(n)}
=
\rho_{\rm crit}^{(0)}
\pm
\Delta\rho_n,
\end{align}
where $\rho_{\rm crit}^{(0)}
\simeq
0.41\,\rho_{\rm Pl}$
is the standard critical density of LQC and $\Delta\rho_n$ denotes the correction generated by the $n$-th fermionic mode. The sign of $\Delta\rho_n$ depends on whether the corresponding mode occupies the vacuum or excited sector.

These corrections should not be interpreted as indicating that the bounce itself becomes fundamentally asymmetric. Rather, different fermionic occupation states define different effective Hamiltonians and therefore different effective cosmological evolutions. Two universes with identical initial quantum geometries but different fermionic quantum states generally follow different relational trajectories while remaining solutions of the same underlying quantum theory.

To characterize these differences quantitatively, we introduce the relational asymmetry function
\begin{equation}
\mathcal A_V(T)
=
\frac{V(T)-V(-T)}
{V(T)+V(-T)}.
\end{equation}
This quantity compares the volume at opposite values of the relational clock. It vanishes identically for a perfectly symmetric bounce,
\begin{equation}
V(T)=V(-T),
\end{equation}
and therefore provides a direct measure of the deviation from exact bounce symmetry induced by fermionic backreaction.

The local behavior of the bounce can be analyzed by expanding the volume around the bounce time $T=0$,
\begin{equation}
V(T)
=
V_b
+
\sigma_1 T^2
+
\sigma_2 T^3
+
\mathcal O(T^4),
\end{equation}
where $V_b$ is the minimum volume attained at the quantum bounce. The coefficient $\sigma_1>0$ determines the local curvature of the bounce, whereas the cubic coefficient $\sigma_2$ represents the leading odd contribution to the expansion. Since odd powers reverse sign under $T\rightarrow -T$, an exactly symmetric bounce requires
\begin{equation}
\sigma_2=0.
\end{equation}
Substituting this expansion into the definition of $\mathcal A_V(T)$ yields
\begin{equation}
\mathcal A_V(T)
=
\frac{\sigma_2}{V_b}T^3
+
\mathcal O(T^5).
\end{equation}
Hence, the coefficient $\sigma_2$ provides a natural quantitative measure of the leading departure from an exactly symmetric bounce.

Within effective theory, $\sigma_2$ is determined entirely by the effective Hamiltonian
\begin{align}
\hat H_{\rm eff}
=
\hat H_o
+
\langle\hat H_n^{(T)}\rangle,
\end{align}
which governs the relational evolution of the quantum geometry. Using Hamilton's equations,
\begin{align}
\dot V
=
\{V,H_{\rm eff}\},
\end{align}
higher-order derivatives are generated recursively through nested Poisson brackets,
\begin{align}
\ddot V
&=
\{\dot V,H_{\rm eff}\},
\\
V^{(3)}
&=
\{\ddot V,H_{\rm eff}\}.
\end{align}
Evaluating the third derivative at the bounce gives
\begin{align}
\sigma_2
=
\frac16
\left.
V^{(3)}
\right|_{T=0}
=
\frac16
\left.
\Bigl\{
\bigl\{
\{V,H_{\rm eff}\},
H_{\rm eff}
\bigr\},
H_{\rm eff}
\Bigr\}
\right|_{T=0}.
\end{align}
This expression establishes a direct connection between the microscopic fermionic backreaction and the macroscopic properties of the bounce. Because the effective Hamiltonian contains the expectation value of the fermionic Hamiltonian, different fermionic occupation states generally lead to different values of $\sigma_2$. The asymmetry coefficient therefore defines a relational observable that quantifies how the quantum state of the fermionic sector influences the effective cosmological evolution.

This construction suggests a useful diagnostic for fermionic backreaction in LQC. Instead of classifying the bounce simply as symmetric or asymmetric, one can compare different quantum states through the observable $\sigma_2$, which measures the leading odd deformation of the effective bounce while remaining fully consistent with the time-reversal invariance of the underlying quantum dynamics.

\subsection{Late-time universe: Emergent cosmological constant}

The analysis of the previous sections shows that fermionic backreaction modifies the effective dynamics of the quantum geometry. We now investigate the behavior of these corrections in the opposite regime, namely the large-volume semiclassical phase well after the quantum bounce. Our goal is to show that, for massive fermions, the backreaction approaches a constant energy density at late times and therefore behaves as an effective cosmological constant. To make this statement precise, we first derive the physical energy density associated with a single fermionic mode and then study its asymptotic behavior in the semiclassical limit.

The natural definition of the physical energy density with respect to the relational time $T$ is
\begin{equation}
\rho_n(T)
=
\frac{\langle \hat H_n^{(T)}\rangle}
{\langle\hat V(T)\rangle},
\label{eq:rho_def}
\end{equation}
where $\hat H_n^{(T)}$ is the physical Hamiltonian of the $n$-th fermionic mode and $\hat V(T)$ is the physical volume operator. After deparametrization with respect to the scalar-field clock, $T$ becomes the physical evolution parameter and $\hat H_n^{(T)}$ generates the relational dynamics. Eq.~\eqref{eq:rho_def} is therefore the natural relational analogue of the classical definition $\rho=E/V$, with both the energy and the volume promoted to physical Dirac observables evaluated at the same relational time.

To evaluate Eq.~\eqref{eq:rho_def}, we consider the semiclassical regime 
\begin{equation} 
V\gg V_b.\nonumber
\end{equation} 
In this limit the mass contribution dominates the fermionic spectrum, since the volume term proportional to $\hat V^2$ grows faster than the kinetic contribution proportional to $\hat V^{4/3}$. Expanding the operator $\hat w_n$ defined in Eq.~\eqref{eq:fermion_spectrum} in powers of $\hat V^{-2/3}$ therefore yields 
\begin{align} 
\hat w_n &= \sqrt{ \frac{\lambda_n^2}{\ell^4}\hat V^{\frac{4}{3}} + \frac{m^2}{\ell^6}\hat V^2 } \nonumber\\ 
&= \frac{m}{\ell^3}\hat V \left[ 1+ \frac{\lambda_n^2\ell^2}{2m^2}\hat V^{-\frac{2}{3}} + \mathcal O(\hat V^{-\frac{4}{3}}) \right]. 
\label{eq:wn} 
\end{align}
The leading contribution is therefore simply \begin{equation} \hat w_n \simeq \frac{m}{\ell^3}\hat V, \end{equation} showing that, at sufficiently large volumes, the physical energy of a massive fermionic mode grows linearly with the physical volume. Consequently, once the energy is divided by the volume to form the energy density, the leading contribution becomes independent of the cosmological expansion.
Using the definition of the relational Hamiltonian, \begin{equation} 
\hat H_n^{(T)} = \hat H_o^{-1/2} \hat w_n \hat H_o^{-1/2}, \end{equation} 
the asymptotic Hamiltonian becomes 
\begin{equation} \hat H_n^{(T)} \simeq m\, \hat H_o^{-1/2} \hat V \hat H_o^{-1/2}. 
\label{eq:HamiltonianCC}
\end{equation}

For sharply peaked semiclassical background states, quantum correlations between the background operators are suppressed. To leading order one may therefore factorize the expectation value as \begin{equation} 
\big\langle \hat H_o^{-1/2} \hat V \hat H_o^{-1/2} \big\rangle\, \simeq\,  \langle\hat H_o^{-1}\rangle \langle\hat V\rangle, 
\end{equation} 
which is the standard approximation for semiclassical states with negligible connected correlations.
Substituting this approximation into the expectation value of the Hamiltonian gives 
\begin{eqnarray} 
\langle\hat H_n^{(T)}\rangle \simeq m\, \langle\hat H_o^{-1}\rangle \langle\hat V\rangle, 
\end{eqnarray} 
and therefore the corresponding physical energy density becomes 
\begin{eqnarray} 
\rho_n \simeq m\, \langle\hat H_o^{-1}\rangle. 
\end{eqnarray}
An important feature of this result is that the leading contribution is independent of the physical volume. Since the background Hamiltonian $\hat H_o$ generates the relational evolution, its expectation value remains constant for semiclassical states evolving under the effective dynamics. Consequently, 
\begin{equation} 
\rho_n\, \simeq\, \mathrm{const.}, 
\end{equation} 
up to corrections of order $\mathcal O(V^{1/3})$ arising from the subleading terms in Eq.~\eqref{eq:wn}.

A constant energy density is the defining property of a cosmological constant in an FLRW universe. Indeed, an effective fluid with  $\rho=\mathrm{const.}$
satisfies the equation of state 
$p=-\rho$, 
and therefore contributes to the Einstein equations exactly as a cosmological constant. The leading late-time fermionic backreaction can therefore be identified with the effective vacuum energy 
\begin{equation} 
\Lambda_{\rm eff} = 8\pi G\, m\, \langle\hat H_o^{-1}\rangle. 
\label{eq:Lambda_eff} 
\end{equation}
This equation provides an explicit derivation of the emergent cosmological constant within the present relational framework. The result follows directly from the asymptotic behavior of the physical Hamiltonian and does not require introducing an independent vacuum-energy term by hand.

To estimate the magnitude of this contribution, we use the classical relation at the quantum bounce, 
\begin{align} 
P_T = V_b \sqrt{2\rho_{\rm crit}^{(0)}}, 
\end{align} 
together with the semiclassical approximation \begin{equation} 
\langle\hat H_o^{-1}\rangle \simeq P_T^{-1}. 
\end{equation} 
This gives 
\begin{align} 
\rho_n \simeq \frac{m} {V_b\sqrt{2\rho_{\rm crit}^{(0)}}}. \end{align} 
For a representative neutrino mass, 
\begin{equation} 
m_\nu \simeq 0.05~{\rm eV}\, \simeq\,  4\times10^{-30}m_{\rm Pl}, 
\end{equation} 
and a typical semiclassical bounce volume 
\begin{equation} 
V_b \sim 10^{8} \text{--} 10^{10}\, V_{\rm Pl}, 
\end{equation} 
one finds 
\begin{eqnarray} 
\rho_n \sim 10^{-38}\text{--}10^{-40}\, \rho_{\rm Pl}. 
\end{eqnarray}
Although this value is extremely small compared with the Planck density, it still exceeds the observed dark-energy density, 
\begin{equation} 
\rho_\Lambda^{\rm obs} \simeq 10^{-123}\rho_{\rm Pl}, \end{equation} 
by approximately $10^{83}$--$10^{85}$ orders of magnitude. The present mechanism therefore naturally generates a cosmological-constant-like contribution, but its magnitude is much larger than the observed value for the semiclassical LQC states considered here.

Conversely, reproducing the observed dark-energy density with a neutrino mass of order $0.05\,{\rm eV}$ would require an enormous bounce volume, \begin{equation} V_b \sim 10^{94}V_{\rm Pl}, \end{equation} which lies many orders of magnitude above the values usually encountered in semiclassical LQC.

Despite this quantitative mismatch, Eq.~\eqref{eq:Lambda_eff} reveals an important conceptual feature of the present construction. Unlike the conventional semiclassical treatment of quantum fields on a fixed classical spacetime, the effective vacuum energy is not determined solely by the particle spectrum. Instead, it depends explicitly on the expectation value of the quantum gravitational Hamiltonian, 
\begin{equation} 
\Lambda_{\rm eff} \propto m\, \langle\hat H_o^{-1}\rangle, \end{equation} 
which is evaluated in the quantum state describing the homogeneous geometry. Consequently, identical fermionic matter content may generate different effective cosmological constants if the underlying quantum states of geometry differ.

The origin of this state dependence can be traced directly to the relational construction of the physical Hamiltonian. After deparametrization with respect to the scalar-field clock, the physical fermionic Hamiltonian is not simply the expectation value of a matter operator but instead takes the form (\ref{eq:HamiltonianCC})  so that the quantum geometry enters explicitly through the operator $\hat H_o^{-1}$. The effective vacuum energy therefore characterizes the coupled matter--geometry quantum state rather than the matter sector alone.

In this sense, the cosmological constant emerging from fermionic backreaction is a genuinely relational observable. Its value is determined jointly by the fermionic mass spectrum and the quantum state of the background geometry, or equivalently by the gravitational Dirac observable generating the relational evolution. This feature appears to be specific to the relational quantization of LQC and has no direct analogue in the conventional semiclassical treatment of quantum fields on a fixed classical spacetime. It therefore suggests a novel perspective on the origin of vacuum energy, in which the observed cosmological constant may depend not only on particle physics but also on the quantum state of the Universe itself.

Finally, it is instructive to compare this behavior with the massless case. Setting $m=0$ in Eq.~\eqref{eq:fermion_spectrum} gives 
\begin{align} 
\hat w_n = \frac{|\lambda_n|}{\ell^2} \hat V^{\frac{2}{3}}, 
\label{eq:wn_massless} 
\end{align} 
so that the corresponding energy density scales as \begin{equation} 
\rho_n^{\rm (massless)}\, \propto\, \big\langle \hat V^{-\frac{1}{3}} \big\rangle\, \propto\,  a^{-1}, 
\label{eq:rho_massless} 
\end{equation} 
where we have used the semiclassical relation $V\propto a^3$.
Unlike the massive contribution, this energy density continues to dilute as the universe expands. It therefore cannot behave as vacuum energy and does not contribute to late-time accelerated expansion. We conclude that a nonvanishing fermion mass is an essential ingredient for the emergence of an effective cosmological constant within the present framework.

\section{Conclusion and outlook}
\label{sec:conclusion}

In this work, we have presented a unified Hamiltonian description of Dirac fermions propagating on LQC backgrounds. Building upon the dressed-metric formulation of LQC, we reviewed the effective dynamics of fermionic fields in the test-field approximation and then discussed how this framework can be extended to incorporate fermionic backreaction through a Born--Oppenheimer treatment. The resulting picture provides a coherent description of the interplay between quantum geometry and fermionic matter, illustrating both how the underlying quantum geometry influences fermionic propagation and how the fermionic sector modifies the effective spacetime experienced by matter.

The principal results discussed in this work may be summarized as follows.
\begin{enumerate} 
\item \textit{Dressed geometry for Dirac fields.} Within the test-field approximation, fermionic modes propagate on an effective dressed spacetime determined by expectation values of quantum geometric operators. Unlike scalar fields, the Dirac Hamiltonian depends on two independent geometric operators associated with the kinetic and mass terms. Consequently, massive fermions are sensitive to modifications of both the temporal and spatial components of the effective metric, whereas massless fermions retain their conformal behavior and are affected only through a redefinition of the conformal time.

\item \textit{Fermionic backreaction and rainbow metrics.} Extending the dressed-metric framework to include backreaction, we discussed a Born--Oppenheimer description in which the quantum geometry becomes dressed by the energy of individual fermionic modes. This naturally leads to mode-dependent effective metrics, providing a realization of quantum-gravity rainbow geometries. Owing to the finite-dimensional Hilbert space associated with each fermionic mode, the backreaction possesses only two physical channels---vacuum and pair-excited states---making the resulting framework considerably simpler than its bosonic counterpart while retaining nontrivial physical consequences. 

\item \textit{Effects on the quantum bounce.} Fermionic backreaction modifies the effective Hamiltonian governing the background dynamics and therefore changes the effective conditions under which the quantum bounce occurs. The resulting corrections depend on the occupation state of the fermionic sector, leading to quantum-state-dependent critical densities and effective bounce trajectories. The underlying quantum dynamics remain invariant under time reversal; the apparent asymmetries discussed in the effective evolution arise from the state dependence of the matter--geometry coupling rather than from a fundamental breaking of time-reversal symmetry.

\item \textit{Late-time effective vacuum energy.} In the semiclassical large-volume regime, we showed that the backreaction of massive fermions approaches a constant energy density and can therefore be interpreted as an emergent cosmological constant. An important feature of this result is that the effective vacuum energy depends not only on the fermionic mass spectrum but also on the quantum state of the background geometry through the expectation value of the inverse background Hamiltonian. This provides a genuinely relational interpretation of vacuum energy within LQC. 

\item \textit{Distinctive role of fermions.} Compared with bosonic fields, fermions exhibit several qualitative differences that make their cosmological backreaction particularly interesting. The Pauli exclusion principle bounds the occupation number of each mode, resulting in finite backreaction strengths and a controlled perturbative treatment. At the same time, this bounded structure gives rise to rich phenomenological effects, including state-dependent dressed geometries, rainbow metrics, and modifications of the effective bounce dynamics. \end{enumerate}

Although much of the formalism presented here builds upon established developments in LQC and quantum field theory on quantum spacetimes, bringing these ingredients together provides a unified framework for understanding the role of Dirac fermions in quantum cosmology. In particular, the discussion highlights how the dressed-metric formalism, fermionic backreaction, rainbow geometries, and the relational interpretation of effective vacuum energy are naturally connected within a single Hamiltonian description.

The present analysis is subject to several simplifying assumptions. The Born--Oppenheimer approximation neglects higher-order non-adiabatic couplings, the backreaction is treated perturbatively, and interactions among fermionic modes are ignored. Extending the formalism beyond these approximations remains an important direction for future work. Several directions for future work remain open. 
An important next step is to perform numerical simulations of the coupled matter--geometry system beyond the perturbative regime in order to quantify the full effects of fermionic backreaction throughout the quantum bounce. It would also be interesting to extend the present framework to anisotropic and inhomogeneous cosmological models, where additional gravitational degrees of freedom may produce qualitatively new fermionic phenomena. Another natural extension is the inclusion of gauge interactions, allowing the study of fermionic matter in more realistic early-universe plasma environments. Finally, the observational consequences of fermionic dressed and rainbow metrics deserve further investigation, particularly their possible imprints on primordial particle production, the cosmic neutrino background, and the effective dark-energy sector.

More broadly, the framework reviewed and extended in this work illustrates that fermions provide a distinctive probe of quantum spacetime. Their spinorial nature, finite-dimensional mode structure, and bounded backreaction lead to features that differ substantially from those of bosonic fields while remaining amenable to analytical treatment. We hope that this synthesis serves as a useful reference for future studies of fermionic quantum fields in LQC and stimulates further investigations into the phenomenology of matter--geometry interactions in the early Universe.

\backmatter

\bmhead{Acknowledgements}
This work has greatly benefited from numerous inspiring discussions with the late Jurek Lewandowski. We are deeply grateful for his guidance and inspiration. We dedicate this paper to his memory in honor of an exceptional scientist and a cherished mentor.
Y.T. acknowledges support from the Iranian Bonyad-e Melli-e Nokhbegan (BMN) via the Kazemi-Ashtiani grant. This work was conducted within the framework of COST Actions CA18108 ``Quantum Gravity Phenomenology in the Multi-Messenger Approach,'' CA23130 ``Bridging High and Low Energies in Search of Quantum Gravity,'' and CA23115 ``Relativistic Quantum Information.''


\end{document}